\documentclass{camera}
\usepackage{graphicx}  % uncomment this if your want to insert figures
\usepackage[latin1]{inputenc}

\newcommand{\nuc}[2]{\ensuremath{^{#1}}#2}

\newcommand{\AM}{$A\,$MeV}

\begin{document}

%%%%%%%%%%%%%%%%%%%%%%%%%%%%%%%%%%%%%%%%%%%%%%%%%%%%%%%%
% The title, only the first letter capitalized; if you want to split it in
% two or more lines, put a \\ macro at each line break
% example: 
%   \title{Title: first line\\ second line}
%
\title{Nuclear thermodynamics with isospin degree of freedom:\\ from
first results to future possibilities with A and Z identification arrays}

%%%%%%%%%%%%%%%%%%%%%%%%%%%%%%%%%%%%%%%%%%%%%%%%%%%%%%%%
% The author(s), separated by commas; do not put a
% comma before the last author, use instead the \and
% macro which produces a normal ``and'' in the
% caps/small caps context
%
\author{ B.~Borderie}

%%%%%%%%%%%%%%%%%%%%%%%%%%%%%%%%%%%%%%%%%%%%%%%%%%%%%%%%
%
\organization{Institut de Physique Nucl\'eaire, CNRS/IN2P3, Univ.
Paris-Sud, Universit\'e Paris-Saclay, 91406 Orsay, France}

\maketitle

\begin{abstract}
Nuclear thermodynamics studies evidenced the existence of a first
order phase transition, namely of the liquid-gas type,
without paying attention to the isospin degree of freedom.
On the other hand, only few results with the introduction of the isospin variable
have been so far obtained. Moreover above all a key question remains.
It concerns the origin of the dynamics of the phase transition: spinodal
instabilities or not with possible consequences and new signatures related to the
introduction of the isospin variable. 
\end{abstract}

\section{Introduction}
 During the last decades nuclear thermodynamics was widely studied through
 heavy-ion collisions at intermediate and relativistic energies and
 through hadron-nucleus collisions. With such collisions,
 depending on impact parameter, a nucleus (or a nuclear system) can be
 heated, compressed, diluted. These systems are expected to undergo a 
 liquid-gas type phase transition that manifests
 through nuclear multifragmentation, This theoretical expectation is
 due to the similarity between the nuclear
 interaction and the Van der Waals forces acting in classical
 fluids~\cite{Bor08}. 
 However a nucleus (or a nuclear system) is a finite system which shows
 specific behaviours in the transition
 region. Most of the predicted specific signals were experimentally
 evidenced without paying attention to the isospin degree of freedom.
 Some are a direct consequence of the abnormal curvature of the
 entropy in the coexistence region which is no more additive due to
 surface (finite size) effects~\cite{Cho04,Bor08}. The answer of a key point is still
 pending, it concerns the dynamics of the transition. And the question
 is: is the phase separation produced by spinodal instabilities or
 produced at equilibrium? At present there is an indication that multifragmentation
 may be induced by  spinodal instabilities but the confidence level of
 the fossil signature is not sufficient (3-4 $\sigma$)~\cite{I31-Bor01,I40-Tab03}.
 Such instabilities
 may happen when the system evolves through the
 mechanically unstable spinodal region of the phase diagram, located at densities
 $\rho\leq\rho_0$ and temperature below the critical temperature. Such conditions are
 well explored in central and mid-peripheral nuclear collisions around the Fermi energy.
 If spinodal instabilities can be experimentally confirmed, new
 signatures are theoretically predicted in relation with the
 introduction of the isospin degree of freedom~\cite{Bar98,Col02,Duc07}, which can be
 experimentally studied.
  
 In this paper, after a brief review of the nuclear phase transition
 signals without paying attention to isospin, we shall 
 discuss first experimental results showing the influence of the
 isospin degree of freedom. Then the key question of the dynamics of
 the transition and the related signatures in relation with isospin
 will be discussed.
   
\section{Direct signatures of first order phase transition in finite nuclear systems}
For finite systems the entropy per particle at equilibrium $S(E)/A$ in the
coexistence region shows a convexity because the entropy of surfaces
which separate the two phases does not scale with $A$.
This behaviour induces specific direct signatures of the phase transition.
Entropy convexity is necessarily accompanied by 
a negative heat capacity in the coexitence zone. This direct signature
was early observed (with a microcanonical sampling) for
 35~\AM{} Au+Au  semi-peripheral collisions~\cite{MDA00} and confirmed
 by the INDRA collaboration for 32-50~\AM{} Xe+Sn central
 collisions~\cite{I46-Bor02}. More
 recently another direct signature associated to convexity was observed:
the bimodal distribution of an order parameter like 
the charge of the largest fragment (Z$_{max}$) of the multifragmentation
partitions. Bimodality was observed (with a canonical sampling) 
in 60-100~\AM{} Au+Au semi-peripheral
collisions, allowing moreover to estimate the latent heat for nuclei close to
gold around 8~MeV and to set the appearance of the pure gas phase above
8.5-10.4~MeV~\cite{I72-Bon09}.

\begin{figure}
\begin{center}
\includegraphics[width=0.48\textwidth]{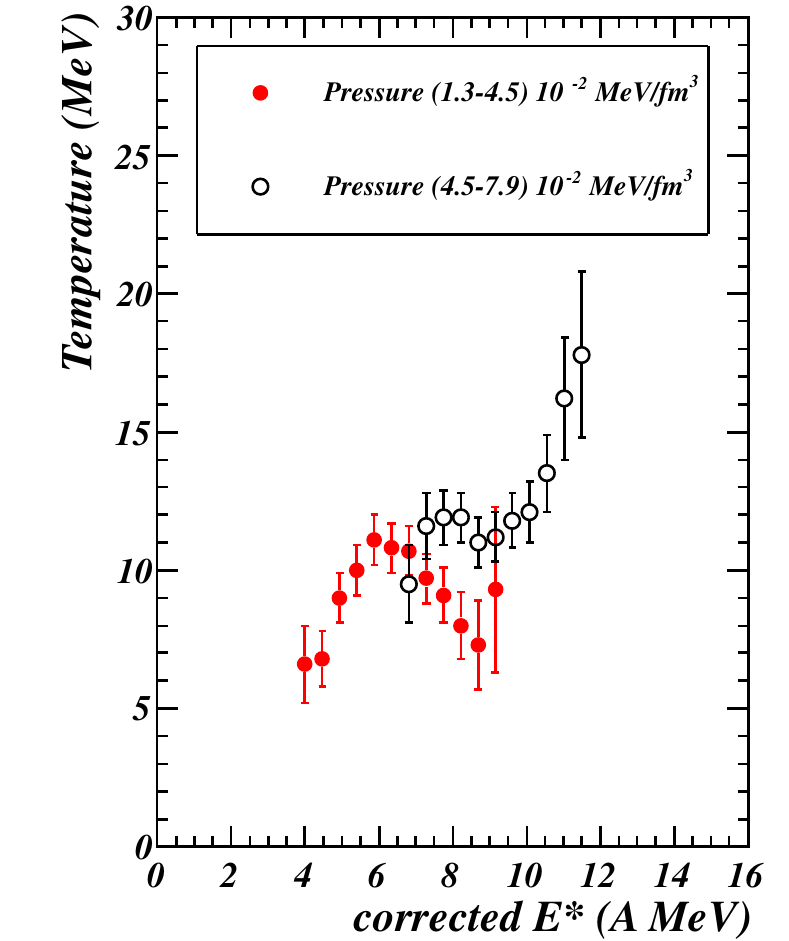}
\includegraphics[width=0.48\textwidth]{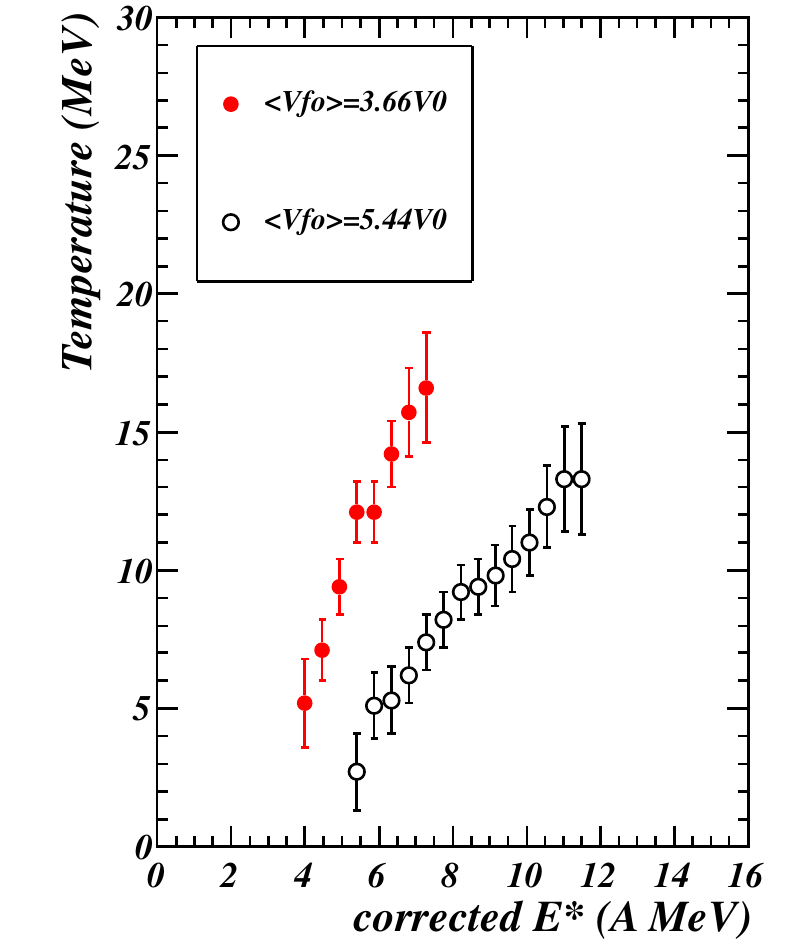}
\end{center}
\caption{ Caloric curves for selected ranges of pressure (left) and
constrained at average volumes (right). Error bars include statistical
and systematic errors. From~\cite{I79-Bor13}.}
\label{fig:BB_CC}
\end{figure}

For caloric curves their shape depends on the
path followed by the system in the microcanonical equation of state
surface, and a backbending
(direct signature) can only be observed for a transition at constant
pressure~\cite{Cho00}. This was evidenced very recently for central
32-50~\AM{} Xe+Sn collisions, thanks to a simulation
based on experimental data~\cite{I66-Pia08} in which
a quantal temperature was calculated from the
momentum fluctuations of protons present at freeze-out~\cite{Zhe11}.  
Pressure and volume-constrained caloric curves could be built and the expected
behaviours were observed:
a backbending for selected ranges of
pressure and a monotonous increase at constant average volume~\cite{I79-Bor13}
(see fig.~\ref{fig:BB_CC}). 

%%%%%%%%%%%%%%%%%%%%% fig.1 %%%%%%%%%%%%%%%%%%%%%%%%%%%%%%% 
%%%%%%%%%%%%%%%%%%%%%%%%%%%%%%%%%%%%%%%%%%%%%%%%%%%%%%%%%%%%%%%%%%%%%%%%%%%%%%

\section{First results with isospin degree of freedom}
 In this section we will briefly review some first studies on phase
 transition signals with the introduction of the isospin variable that
 were experimentally observed in the past years.

   \subsection{The caloric curves}
   A few experiments studied the effect of isospin on caloric curves, considering 
   semi-peripheral collisions. Sfienti \emph{et al.} for 600~\AM{}
   (\nuc{124}{Sn},\nuc{124}{La},\nuc{107}{Sn})+Sn~\cite{Sfi09} and
   Wuenschel \emph{et al.} for 35~\AM{} \nuc{78}{Kr}+\nuc{58}{Ni}, 
   \nuc{86}{Kr}+\nuc{64}{Ni}~\cite{Wue10} found a small
   isospin effect, with slightly higher temperatures for the 
   neutron-richer systems; conversely McIntosh \emph{et al.}, for light 
   quasi-projectiles of known A and Z formed in 35~\AM  \nuc{70}{Zn}+\nuc{70}{Zn}, 
   \nuc{64}{Zn}+\nuc{64}{Zn}, \nuc{64}{Ni}+\nuc{64}{Ni} observed measurable effects, 
   with lower temperatures for neutron-richer nuclei~\cite{McI13}. Note that,
   unlike the ensemble of caloric curves presented in~\cite{Nat02}, none of those
   derived in~\cite{Sfi09,Wue10,McI13} exhibits a plateau. In~\cite{McI13} 
   the temperature linearly increases with energy between 2 and 8~\AM, reaching 
   12 MeV at $E^*$ = 8~\AM, well above the empirical value of $T_{lim}$ for light 
   nuclei estimated in~\cite{Nat02}. 

%%%%%%%%%%%%%%%%%%%%% fig.2 %%%%%%%%%%%%%%%%%%%%%%%%%%%%%%% 
\begin{figure}
\begin{center}
\includegraphics[width=0.95\textwidth]{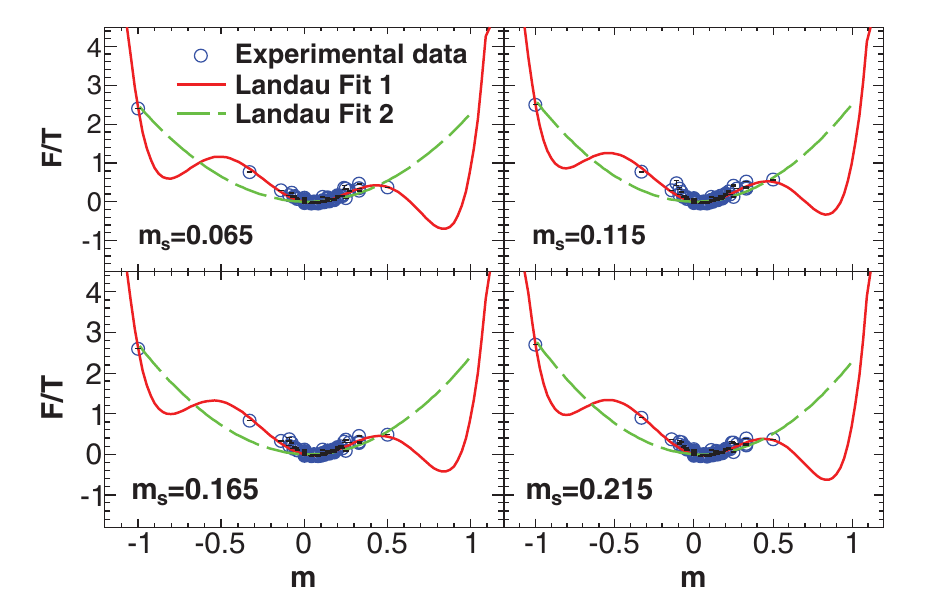}
\end{center}
\caption{
$F$/$T$ as a function of fragment neutron-proton asymmetry $m$ for an
excitation energy of 6.5 MeV per nucleon of the QP. Different panels
correspond to different neutron-proton asymmetry ($m_s$) bins of the QP.
Full lines are fits to the data using the complete Landau free 
energy (first order phase transition)
and dashed lines are fits which refer to single phase systems.
Statistical errors are smaller than the points. From~\cite{Mab13}.}
\label{fig:freeE}
\end{figure}
%%%%%%%%%%%%%%%%%%%%%%%%%%%%%%%%%%%%%%%%%%%%%%%%%%%%%%%%%%%%%%%%%%%%%%%%%%%%%%   
   On the theoretical side, the SMM model, as well as calculations considering a
   nucleus in equilibrium with its vapor, predict an increase of $T_{lim}$ with
   isospin~\cite{Ogu11,Bes89}, whereas in calculations dealing with isolated
   mononuclei, without surrounding vapor, the reverse trend is obtained~\cite{Hoe07}. 
   In all cases the temperature variation with isospin is small.

\subsection{Phase transition and free energy}
    A new signature of first order phase transition was experimentally investigated
    using the Landau free-energy approach~\cite{Mab13}.
    Quasi-projectiles (QP) formed in  35~\AM   \nuc{64}{Zn}+\nuc{64}{Zn}, 
   \nuc{70}{Zn}+\nuc{70}{Zn} and \nuc{64}{Ni}+\nuc{64}{Ni} were
   reconstructed and data sorted in different QP asymmetry
   ($m_s = (N_s-Z_s)/A_s)$ and excitation energy bins in the range
   3-9 MeV per nucleon. According to the modified Fisher model to take into
   account finite size effects, the free energy per nucleon of a
   fragment of mass $A$ normalized to the temperature of the QP, $F/T$, can be
   derived from the fragment yield $Y = y_0 A^{-\tau} e^{(-F/T)A}$;
   $y_0$ is a constant and $\tau$ is a critical exponent. In the
   Laudau approach the free energy of a first order phase transition
   is extended in a power series (sixth order) in the order
   parameter $m = (N_f-Z_f)/A_f)$; $N_f$, $Z_f$ and $A_f$ are the
   neutron, proton and mass numbers of the fragment respectively.
    More details can be found in~\cite{Mab13}.
    In fig.~\ref{fig:freeE} $F/T$ values as a function of $m$ are
    displayed for a QP excitation energy of 6.5 MeV per nucleon and
    for different asymmetry ($m_s$) bins of the QP (different panels).
    Full lines correspond to the true Landau fit with the
    appearance of three minima, which is the signature of a first
    order phase transition. Dashed lines correspond to single
    phase systems.

\section{Dynamics of the transition:\\ spinodal instabilities or not? }
\begin{figure}
\begin{center}
\includegraphics[width=0.95\textwidth]{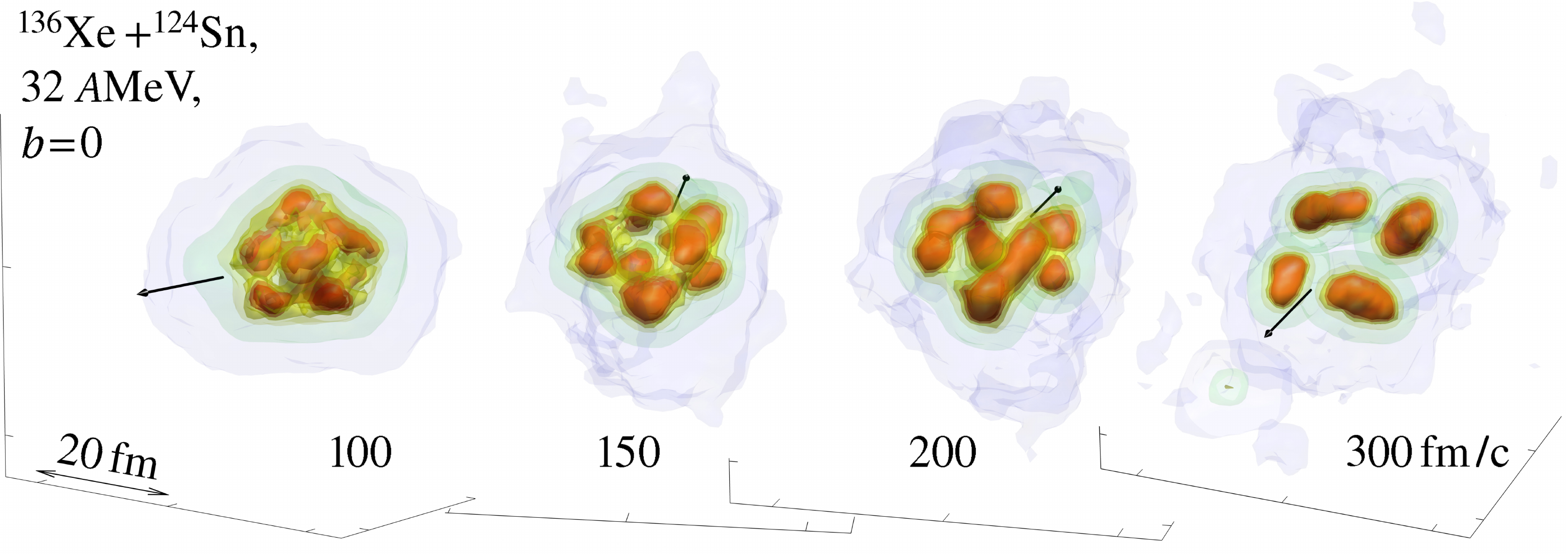}
\end{center}
\caption{
BLOB calculation for central \nuc{136}{Xe} +\nuc{124}{Sn} collisions 
at 32~\AM. Multifragmentation occurs with the spinodal pattern at
150fm/c followed by a partial coalescence at later time. For better
visibility, axis is modified for each time; arrows indicate the
reaction axis. From~\cite{Nap15}.}
\label{fig:BLOB}
\end{figure}

In infinite nuclear matter the signature of spinodal instabilities is the formation
of equal size fragments. The most unstable modes correspond to
wavelengths lying around $\lambda \approx$ 10 fm and the associated
characteristic times are almost identical, around 30-50 fm/c,
depending on density ($\rho_0$/2 - $\rho_0$/8) and 
temperature (0-9 MeV)~\cite{Bor08}. A direct consequence
of the dispersion relation is the production of ``primitive'' fragments
with size $\lambda$/2 $\approx$ 5 fm which correspond to $Z$ $\approx$ 8 .
However this simple picture is
rather academic. A more realistic picture can be obtained from
semi-classical calculations of collisions. Fig.~\ref{fig:BLOB} shows
such results for recent calculations for collisions between
\nuc{136}{Xe} and \nuc{124}{Sn} at 32~\AM~\cite{Nap15} for central
collisions and their evolution with time. In such simulations (BLOB)
fluctuations are introduced in
full phase space from inducing nucleon-nucleon collisions~\cite{Nap13}.
One note that despite finite size effects (equally probable modes with
beating) a large number of ``primitive'' fragments formed at 150fm/c
(around three times the characteristic time) are of comparable sizes
in the region from Ne to O. Then coalescence occurs, which largely
destroys the ``primitive'' composition and which well
confirms that the final extra production of equal-sized fragments is a fossil
signature. It is the reason why such signature
is difficult to observe if the spinodal decomposition occurs.

%%%%%%%%%%%%%%%%%%%%% fig.3 %%%%%%%%%%%%%%%%%%%%%%%%%%%%%%% 
%%%%%%%%%%%%%%%%%%%%%%%%%%%%%%%%%%%%%%%%%%%%%%%%%%%%%%%%%%%%%%%%%%%%%%%%%%%%%%
As said in the introduction, a definitive conclusion concerning the
dynamics of the transition was not obtained experimentally in relation
with a too low confidence level (3-4 $\sigma$) for the extra production
of equal-sized fragment partitions~\cite{I31-Bor01,I40-Tab03}.
Work is in progress in the INDRA collaboration to
give a final conclusion from experiments realized with
much higher statistics. Various reactions have been used, 
 \nuc{124,136}{Xe}+\nuc{112,124}{Sn} at 32 and 45~\AM, to produce quasi
fusion (QF) systems. One can also expect possible information related to
the influence of isospin on spinodal instabilities from QF events formed
from extreme reactions (difference of twenty four neutrons).

\subsection{Spinodal region and isospin degree of freedom}
 First of all, asymmetric nuclear matter at subsaturation densities is
 shown to present only one type of instability, which means a unique
 spinodal region~\cite{Mar03}. The associated order parameter is
 dominated by the isoscalar density  which implies a transition of the
 liquid-gas type. 
 The spinodal zone is also predicted to shrink for increasing isospin asymmetry,
 reducing both the critical density and temperature~\cite{Bar98}.
 For finite size systems like nuclei the direct consequences are the
 following: heavier systems have a larger instability region than the
 lighter ones and more asymmetric systems/nuclei are less unstable
 (see fig.~\ref{fig:spino_I_nuclei}), which
 means a reduced spinodal zone compared to symmetric systems~\cite{Col02}.
 If spinodal instabilities are at the origin of the phase 
 transition any change of the spinodal region in the
 phase diagram should affect the signals previously discussed.

 \subsection{Spinodal instabilities versus equilibrium as dynamics of
 the transition}
%%%%%%%%%%%%%%%%%%%%%%%%%% fig. 4 et 5 %%%%%%%%%%%%%%%%%%%%%%%%%%%%%%%%%%
   \begin{figure}
       \begin{minipage}[t]{0.56\textwidth}
	  \includegraphics[width=\textwidth]{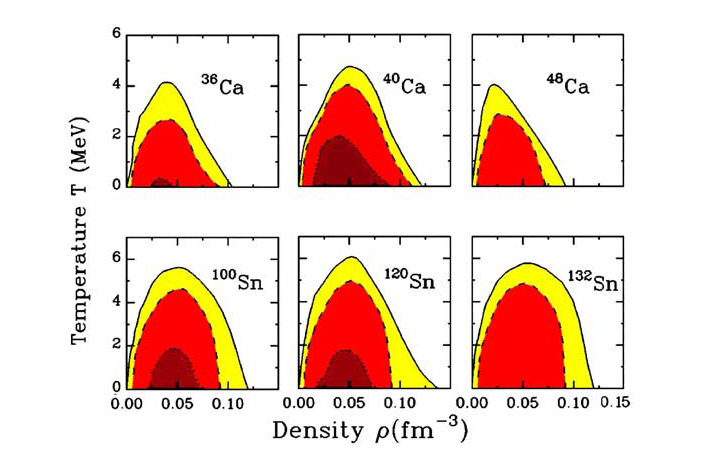}
	  \caption{Border of the instability region. Boundaries of the
	  instability region (solid lines) associated with $L$ = 3
	  collective modes in Ca and  Sn isotopes. Phase points
	  having the same growth time equal to either 100fm/c (dashed)
	  or 50fm/c (dots) are also delineated. From~\cite{Col02}.} \label{fig:spino_I_nuclei}
	\end{minipage}%
	\hspace*{0.02\textwidth}%
	\begin{minipage}[t]{0.42\textwidth}
	  \includegraphics[width=\textwidth]{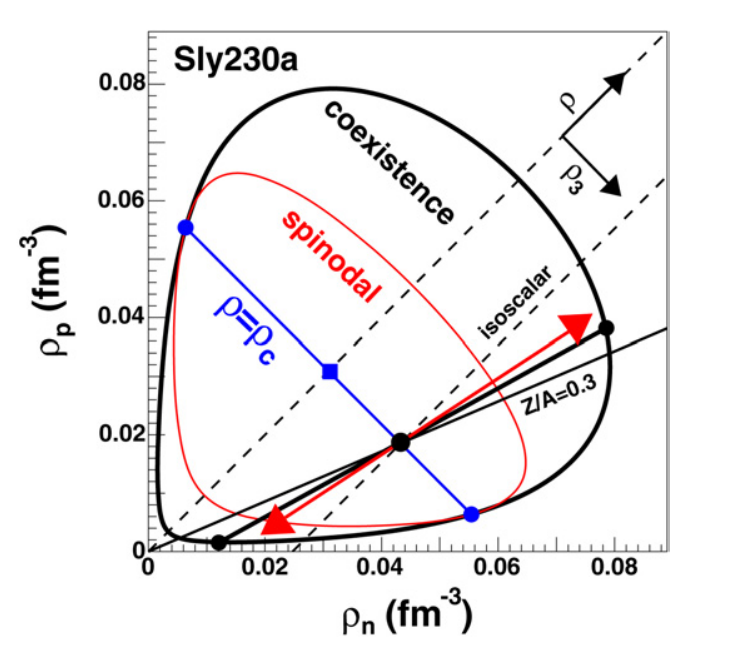}
	  \caption{Coexistence and spinodal regions in the
	  proton-neutron density plane at $T$ = 10 MeV. Illustration
	  of phase separation inside the instability region for matter with a proton
	  fraction $Z$/$A$ = 0.3 (see text). From~\cite{Duc07}.} \label{fig:EqSpino}
	\end{minipage}
   \end{figure}
   %%%%%%%%%%%%%%%%%%%%%%%%%%%%%%%%%%%%%%%%%%%%%%%%%%%%%%%%%%%%%%%%%%%%%%%%%%%%%%%%%
 Considering the asymmetric nuclear matter liquid-gas phase transition
 analyzed in a mean field approach, two different mechanisms of phase
 separation (dynamics of the transition) can be compared: equilibrium
 and spinodal decomposition~\cite{Duc07}. The isospin properties of
 the phases are deduced from the free-energy curvature, which contains
 information on the average isospin of the phases and on the system
 fluctuations. The results are illustrated in fig.~\ref{fig:EqSpino}
 for neutron rich matter with $Z$/$A$ = 0.3 and a temperature $T$ of
 10 MeV. If equilibrium is the origin of phase separation, the system
 will undergo separation according to Gibbs construction. The two
 phases, represented as black dots on the coexistence border do not belong
 to the line of constant proton fraction ($Z$/$A$ = 0.3). The liquid
 fraction is closer to symmetric matter than the gas phase. It is a
 consequence of the symmetry energy minimization in the dense phase.
 This inequal repartition of isospin between the two phases is the
 well-known phenomenon of isospin fractionation.
 One can now study isospin fractionation if phase separation is driven
 by spinodal instabilities. This is the spinodal decomposition and the
 local properties of the constrained entropy curvature determine the
 development into phase separation. The double arrow in fig.~\ref{fig:EqSpino}
 shows the results; the spinodal decomposition leads to a more pronounced
 fractionation than equilibrium, the dense phase getting closer to the
 symmetric matter.
 
 This fact can be a possible new signature of the dynamics of the
 phase transition. However it appears as a very challenging problem.
 For experimentalists large $Z$/$A$ values are required to have enough
 sensitivity. A robust reconstruction of primary fragments is also
 mandatory. Moreover future $A$ and $Z$ identification arrays, like for example
 FAZIA, are absolutly needed for such studies~\cite{Bou14,Sa16}.
 On the theoretical side more realistic calculations involving
 collisions between nuclei are essential.

\section{Conclusions}
The presence of a first order phase transition for hot nuclei (or
nuclear systems) seems well established and is not contradicted by recent
results taking into account the isospin degree of freedom.
A key point still remains: the origin of
the dynamics of the transition which is unknown.
If spinodal decomposition occurs, it is confirmed by recent improved realistic
calculations involving collisions that the extra production of
equal-sized fragments is a fossil signature and it can be difficult to
experimentally evidence such a signal. Works are in progress with high
statistics to possibly give a final answer.
The introduction of the isospin degree of freedom into the
game opens possibilities for new signatures of spinodal instabilities.
A direct one is the reduction of the spinodal region for asymmetric
systems. A second more indirect concerns isospin fractionation. The
spinodal decomposition leads to a more pronounced fractionation than
equilibrium, which is the alternative for the dynamics of the transition.
However this last signature is very difficult to quantitatively obtain
at both levels experimental and theoretical.

%%%%%%%%%%%%%%%%%%%%%%%%%%%%%%%%%%%%%%%%%%%%%%%%%%%%%%%%
% Write the text starting from here and using the usual
% LaTeX commands.
%
%%%%%%%%%%%%%%%%%%%%%%%%%%%%%%%%%%%%%%%%%%%%%%%%%%%%%%%%
% End of the paper

\end{document}